\documentclass[prb,twocolumn,showpacs,superscriptaddress]{revtex4}

\usepackage{graphicx}
\usepackage{amsmath,amssymb}
\usepackage{amsbsy}
\usepackage{bm,subfigure,epsfig}

\begin{document}


\newcommand{\comment}[1]{{\bf (#1)}}
\newcommand{\up}{\uparrow}
\newcommand{\down}{\downarrow}
\renewcommand{\d}{{\rm d}}
\newcommand{\e}{{\rm e}}
\newcommand{\imai}{{\rm i}}
\newcommand{\cs}{\cos\!\frac{\phi}{2}}
\newcommand{\si}{\sin\!\frac{\phi}{2}}
\newcommand{\Eo}{E_{\tilde{\uparrow}}^{{}}}
\newcommand{\En}{E_{\tilde{\downarrow}}^{{}}}
\newcommand{\Ek}{E_k^{{}}}
\newcommand{\Ekp}{E_{k'}^{{}}}
\newcommand{\Eto}{E_2^{{}}}
\newcommand{\To}{T_{\tilde{\uparrow} 0}}
\newcommand{\Tn}{T_{\tilde{\downarrow} 0}}
\newcommand{\Ttoo}{T_{2\tilde{\uparrow}}}
\newcommand{\Tton}{T_{2\tilde{\downarrow}}}
\newcommand{\ket}[1]{{|#1\rangle}}
\newcommand{\upTild}{{\tilde{\uparrow}}}
\newcommand{\downTild}{{\tilde{\downarrow}}}
\newcommand{\ave}[1]{{\langle #1 \rangle}}


\title{Interplay between interference and Coulomb interaction in the
  ferromagnetic Anderson model with applied magnetic field}

\author{Jonas Nyvold Pedersen}
\affiliation{Mathematical Physics, Physics Department, Lund
University, Box 118, SE-22100 Lund, Sweden}
\author{Dan Bohr}
\affiliation{Department of Physics, University of Basel,
Klingelbergstr.~82, CH-4056 Basel, Switzerland}
\author{Andreas Wacker}
\affiliation{Mathematical Physics, Physics Department, Lund
University, Box 118, SE-22100 Lund, Sweden}
\author{Tom\'{a}\v{s} Novotn\'{y}}
\affiliation{Department of Condensed Matter Physics, Faculty of
Mathematics and Physics, Charles University, Ke Karlovu 5, 121 16
Prague, Czech Republic} %
\affiliation{Nano-Science Center and  Niels Bohr Institute,
University of Copenhagen, Universitetsparken 5, 2100 Copenhagen,
Denmark}
\author{Peter Schmitteckert}
\affiliation{Institut f\"ur Nanotechnologie, Forschungszentrum
Karlsruhe, 76021 Karlsruhe, Germany}
\author{Karsten Flensberg}
\affiliation{Nano-Science Center and  Niels Bohr Institute,
University of Copenhagen, Universitetsparken 5, 2100 Copenhagen,
Denmark}

\date{\today}

\begin{abstract}
We study the competition between interference
due to multiple single-particle paths and Coulomb interaction in a
simple model of an Anderson-like impurity with
local-magnetic-field-induced level splitting coupled to
ferromagnetic leads. The model along with its potential
experimental relevance in the field of spintronics serves as a
nontrivial benchmark system where various quantum transport
approaches can be tested and compared. We present results for the
linear conductance obtained by a spin-dependent implementation of
the density matrix renormalization group scheme which are compared
with a mean-field solution as well as a seemingly more advanced
Hubbard-I approximation. We explain why mean-field yields nearly
perfect results, while the more sophisticated Hubbard-I approach
fails, even at a purely conceptual level since it breaks
hermiticity of the related density matrix. Furthermore, we study
finite bias transport through the impurity by the mean-field
approach and recently developed higher-order density matrix
equations. We find that the mean-field solution fails to describe
the plausible results of the higher-order density matrix approach
both quantitatively and qualitatively as it does not capture some
essential features of the current-voltage characteristics such as
negative differential conductance.
\end{abstract} 

\maketitle

\section{Introduction}
When electrons pass through a mesoscopic region, the superposition
of several different single-particle transport paths can lead to
interference, as, e.g., in an Aharonov-Bohm geometry  with quantum
dots embedded in the arms.\cite{YacobyPRL1995, SigristPRL2004} As
the size of the mesoscopic region diminishes, many-particle
effects such as Coulomb blockade become increasingly
important.\cite{Sohn1997} This may change the amplitudes of the
competing transport paths and thereby alter the interference
effect. Eventually, for sufficiently strong many-body interaction,
the single-particle-paths picture breaks down and such systems
should be treated using a true many-body formalism. This problem
of the interplay between interference of several competing paths
and many-body interaction has recently attracted a lot of
attention theoretically in the general quantum-transport context
\cite{BoesePRB2001,KonigPRB2002,GolosovPRB2006,ZitkoPRB2006,MedenPRL2006,
TokuraNJP2007,DinuPRB2007,FranssonPRB2007,KashcheyevsPRB2007,KonikPRL2007,LobosPRB2008,
KoertingPRB2008,WeymannPRB2008,LuCM2008} as well as from  more
specific points of view such as the molecular electronics,
\cite{HettlerPRL2003,CardamoneNL2006,StaffordNa2007,BegemannPRB2008,SolomonJCP2008,SanvitoPRB2008,KeCM2008}
spintronics,
\cite{KoenigPRL2003,RudzinskiPRB2005,WeymannEPJ2005,CohenPRB2007,TrochaPRB2007}
or even full counting statistics\cite{UrbanPRB2008} and
superconducting transport.\cite{OsawaCM2008}

In the case of spintronics, the interference can be achieved
without necessity of a multiply-connected {\em orbital} geometry
due to the possibility of superposition of different {\em purely
spin} amplitudes with the help of either non-collinearly
magnetized leads \cite{FranssonEPL2005, WeymannEPJ2005,
RudzinskiPRB2005} or an additional spin-level splitting
non-collinear with the lead magnetization.
\cite{PedersenPRB2005b,WeymannEPJ2005} Since experiments with
strongly-interacting quantum dots and ferromagnetic contacts have
recently been successfully performed
\cite{MolenPRB2006,LiuNT2007,HauptmannNP2008} the spin
interference effects proposed in previous
work\cite{PedersenPRB2005b} and further elaborated here may be
within experimental reach.

\begin{figure}
  \includegraphics[trim = 0mm 0mm 0mm 0mm, clip, width=0.6\columnwidth]{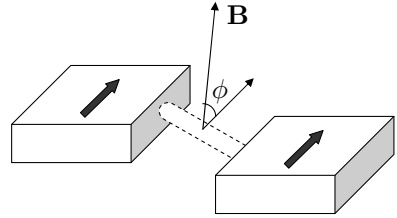}
  \caption{Sketch of our model, where the magnetic field $B$ in the
  central region is tilted by an angle $\phi$ with respect to the magnetization of the
  contacts.}\label{Fig:model}
\end{figure}

\begin{figure}
  \includegraphics[trim = 0mm 0mm 0mm 0mm, clip, width=0.8\columnwidth]{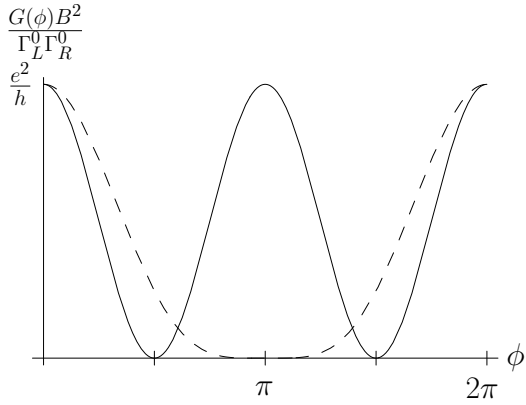}\\
  \caption{The linear conductance of Eq.~(\ref{EqCotunFull}),
  which shows the anti-resonances for non-interacting electrons (full line, $U=0$) and
  the spin valve behavior for strongly interacting electrons (dashed line, $U=100B$) as a function of
  the angle between the
  magnetizations of the leads and the applied magnetic field. }\label{FigConductance}
\end{figure}

Previously, some of us considered a model\cite{PedersenPRB2005b}
consisting of a spin-$1/2$ level coupled to ferromagnetic leads
with the magnetizations being either parallel or antiparallel. In
addition, a magnetic field non-collinear with the spin direction
of the leads was applied (see
Fig.~\ref{Fig:model}).\footnote{Similar models were studied by
others in
Refs.~[\onlinecite{KashcheyevsPRB2007},\onlinecite{WeymannEPJ2005}].}
For this Ferromagnetic Anderson model with an applied magnetic
field $B$ (from now on nicknamed the FAB model) the linear
conductance was obtained in two different regimes: without
interactions on the dot, and in the cotunneling regime. For the
non-interacting case with fully polarized leads, zero temperature,
the bare level on resonance and a parallel lead configuration, the
linear conductance can be calculated, e.g., using non-equilibrium
Green functions (NEGF), and the exact result
is\cite{PedersenPRB2005b} ($h$ is Planck's constant)
\begin{equation}\label{EgGFU0}
G^\mathrm{non-int}=\frac{e^2}{h}\frac{\Gamma^0_L\Gamma^0_R}{B^2}\frac{\cos^2\phi}{1
+\cos^2\phi\left[(\Gamma^0_L+\Gamma^0_R)/2B\right]^2},
\end{equation}
with $\Gamma^0_\alpha$ being the coupling to the leads with
$\alpha=L,R$, $B$ the magnetic field times the magnetic moment of
the level, and $\phi$ the angle between the magnetization in the
leads and the applied magnetic field. The conductance shows
anti-resonance at angles $\phi=\pi/2$ and $\phi=3\pi/2$ due to
destructive interference, see Fig.~\ref{FigConductance}.

Under the same conditions as stated above, the linear conductance
can be obtained in the cotunneling regime
($[\Gamma_L^0+\Gamma_R^0]/B\ll 1$) even in the presence of an
on-site Coulomb interaction $U$, see Eq.~(\ref{HD}), by applying a
scattering formalism: \cite{PedersenPRB2005b}
\begin{equation}\label{EqCotunFull}
G^\mathrm{cotun}=\frac{e^2\Gamma_L^0\Gamma_R^0}{h}
\left[\frac{\cos^2(\phi/2)}{-B}+ \frac{\sin^2(\phi/2)}{B+U}
\right]^2.
\end{equation}
In this regime, the conductance shows a cross-over from the
behavior with anti-resonances at $\phi=\pi/2$ for the
non-interacting case, to a spin-valve effect for
$U\rightarrow\infty$ with $G^\mathrm{cotun}\propto
\cos^4(\phi/2)$. That is the anti-resonances around $\phi=\pi/2$
disappear and the conductance vanishes for $\phi=\pi$ instead (see
Fig.~\ref{FigConductance}).

\begin{figure}
  \includegraphics[trim = 0mm 0mm 0mm 0mm, clip, width=0.6\columnwidth]{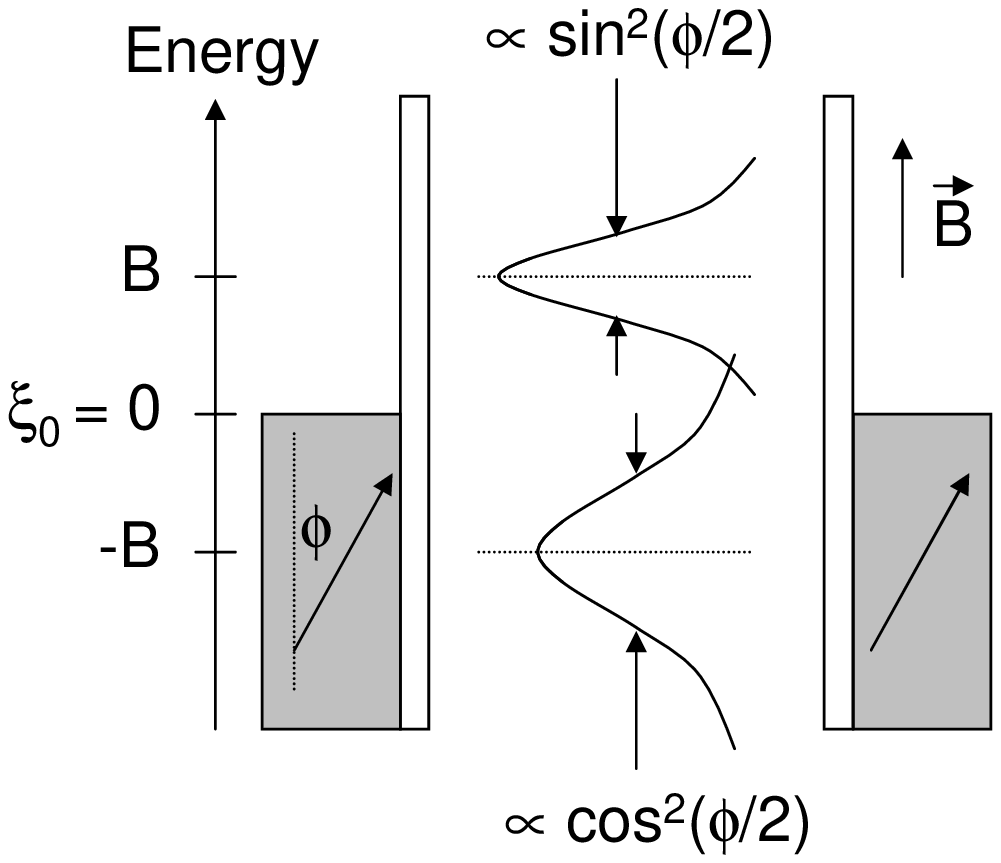}\\
  \caption{
Schematic energy spectrum in the linear conductance regime for the
non-interacting case. The bare resonant level is split due to
magnetic field, and  the angle between the magnetizations of the
leads and the applied magnetic field is denoted by $\phi$. The
widths of the two resonances depend on $\phi$ as $\cos^2(\phi/2)$
and $\sin^2(\phi/2)$, respectively.}\label{FigTunneling}
\end{figure}

A simple physical picture for the situation described above is: In
a basis where the Hamiltonian for the isolated dot is diagonal,
the bare dot level energy is split by the magnetic field, and for
non-interacting electrons the density of states has peaks at the
two single-particle energies at $\mp B$, see
Fig.~\ref{FigTunneling}. The widths of the two peaks depend on the
angle $\phi$, and for fully polarized leads they are proportional
to $\cos^2(\phi/2)$ or $\sin^2(\phi/2)$, respectively. For
$\phi=0$ and $\phi=\pi$ one of the peaks is infinitely narrow and
electrons can only pass through the other level, whereas for
$\phi=\pi/2$ the peaks are equally wide resulting in the sharp
anti-resonances due to interference. So the angular dependence of
the conductance can be understood as interference through
non-degenerate levels, which have widths depending on the angle
between the magnetizations of the leads and the applied magnetic
field. For a large on-site Coulomb interaction some weight of the
density of states is moved away from the single-particle energies
and away from the Fermi level, thereby destroying the
anti-resonances.

The qualitative difference between the interacting and
non-interacting regime is important, as it shows a very crucial
feature in the transport through mesoscopic systems, namely that
it is generally not the single-electron transport paths which
determine the transport, but rather many-electron processes.
Besides from leading to interesting physical effects, it also puts
strong demands on the theoretical transport formalism applied to
such systems, as it should be able to handle both the coherence
and the interactions. It also has to be applicable for
sufficiently low temperature, because otherwise thermal
fluctuations will wash out the interference effect. That makes our
model an excellent benchmark for transport formalisms. Including
interactions, if they are sufficiently strong, is a challenge in
the standard NEGF formalism where all single-particle effects
including the interference are captured exactly. On the other
hand, the density matrix language (generalized master equation;
GME) starting from exact many-body states of the system, thus
including the interaction exactly, faces problems when the
broadening due to the leads comparable with level splitting
(leading to interference effects) is to be incorporated. Thus,
this kind of models poses significant challenges to standard
transport approaches even outside notoriously difficult
strongly-correlated regimes such as the Kondo regime.

Therefore, we use this model for a detailed comparison study of
the performance of different transport formalisms in the
potentially problematic and so far not addressed regime of
broadening comparable to the level splitting
$B\approx\Gamma^0_{\alpha}$ and arbitrarily strong interaction
$U$. In particular, we test the results of higher-order, i.e.
beyond mean-field, decoupling schemes based on NEGF
\cite{RudzinskiPRB2005,KashcheyevsPRB2006,FranssonPRB2007} and/or
many-body-states-based NEGF (Hubbard operator NEGF) approaches.
\cite{FranssonPRL2002,SandalovIJQC2003,FranssonPRB2005,DattaCM2006,FrobrichPhR2006,BergfieldCM2008,GalperinPRB2008}
We find that the Hubbard-I approximation in the framework of
NEGF,\cite{MeirPRL1991,FranssonPRB2005} frequently applied to the
Anderson model with or without ferromagnetic
leads,\cite{RudzinskiPRB2005,FranssonPRB2007,BergfieldCM2008,GalperinPRB2008}
gives unphysical and even mathematically wrong results for the
model considered in this paper. This finding raises serious
questions about the very foundation of the many-body-states-based
NEGF
approaches.\cite{FranssonPRL2002,SandalovIJQC2003,GalperinPRB2008}

\begin{figure}
  \includegraphics[trim = 0mm 0mm 0mm 0mm, clip,width=0.6\columnwidth]{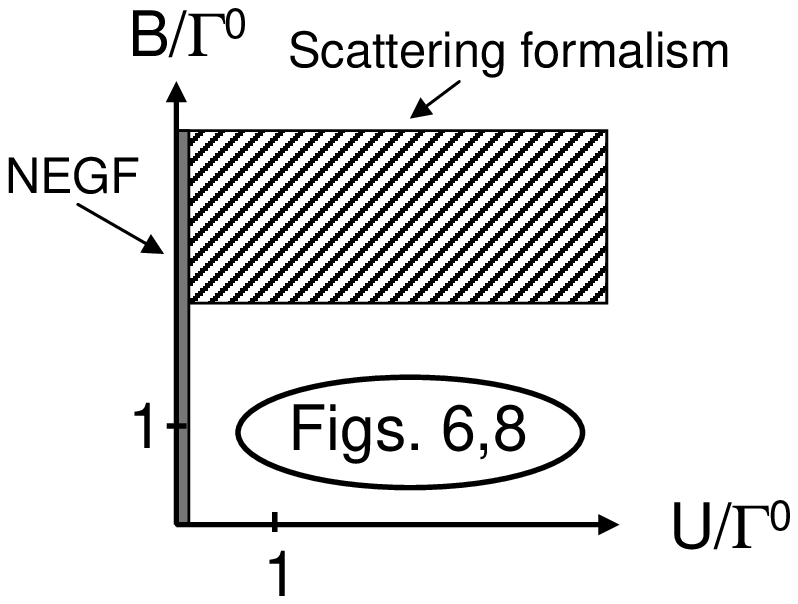}\\
  \caption{Sketch of the different parameter regimes. For vanishing
    interaction $U$, nonequilibrium Green functions
    provide the complete solution of the transport problem, see Eq.~(\ref{EgGFU0}). For large level
    splitting, the scattering formalism allows for a quantitative description of the cotunneling events,
     see Eqs.~(\ref{EqCotunFull}),(\ref{EqCotunPartial}). In this
    work we provide results for the more intricate region of moderate level
    splitting and a finite on-site Coulomb interaction.
    }\label{FigParameters}
\end{figure}

As stated above, so far only the non-interacting and the
cotunneling regime have been considered (see
Fig.~\ref{FigParameters}), and for the latter only in linear
response. In this paper, we calculate the linear conductance at
zero temperature for arbitrary values of the tunneling coupling,
applied magnetic field and on-site Coulomb interaction using a
density matrix renormalization group (DMRG) scheme, see
Sec.~\ref{SecDMRG}. Surprisingly, in the linear response regime
the results obtained using the DMRG scheme can, in certain
situations, be reproduced using Green functions with a mean-field
approximation, which is discussed in Sec.~\ref{SecMF}. The
unexpected failure of the Hubbard-I approximation in the framework
of NEGF\cite{MeirPRL1991,FranssonPRB2005} is analyzed in
Sec.~\ref{SecHubbard1}. In section~\ref{Sec2vN} we extend the
calculations beyond linear response by applying a generalized
master equation formalism\cite{PedersenPRB2005a} which works in a
basis of many-particle states and takes into account higher-order
tunneling processes. In Sec.~\ref{SecFinitebias_Comp} the failure
of the mean-field Green function method for finite bias is
demonstrated.  Finally, we conclude on our findings in
Sec.~\ref{SecConclusion}. Appendix~\ref{AppCotun} contains the
cotunneling expression for less than full polarization of the
leads and off-resonant transport, and App.~\ref{AppNEGF} presents
details of the mean-field Green function calculation.

\section{The model system} \label{SecModel}
The model Hamiltonian of the quantum dot coupled to
magnetic leads is
\begin{equation}  \label{Hmodel}
H=H_{LR}^{{}}+ H_T + H_D,
\end{equation}
where
\begin{equation}  \label{HLR}
H_{LR}^{{}}=\sum_{\alpha=L,R,k\sigma} \xi_{\alpha,k\sigma}^{{}}
c^\dagger_{\alpha,k\sigma} c^{{}}_{\alpha,k\sigma}.
\end{equation}
Here $\sigma=\uparrow/\downarrow$ is the spin of the electrons,
$\alpha$ denotes the left or right electrodes, which are assumed
to be polarized along the $z$-axis (the spin quantization axis),
either parallel or anti-parallel. However, in this paper we only
consider parallel magnetizations of the leads. The quantum dot is
subjected to a magnetic field ${\bf B}$, which is tilted by an
angle $\phi$ with respect to the $z$-axis and lies within the
$xz$-plane. Note that we neglect the negative sign of the electron
charge for simplicity. Thus, the energetically preferred spin
direction is pointing in the direction of {\bf B} throughout this
paper. The dot-Hamiltonian reads ($n_\sigma=d^\dag_\sigma
d_\sigma$)
\begin{equation}  \label{HD}
H_{D}^{{}}=\sum_{\sigma} \xi_{0}^{{}}d^\dagger_\sigma
d^{{}}_\sigma
+Un_{\uparrow}^{{}}n_{\downarrow}^{{}}-\sum_{\sigma\sigma^{\prime}}\,\mu_B\mathbf{B}
\cdot\pmb{\tau}_{\sigma\sigma^{\prime}}^{{}} d^\dagger_\sigma
d^{{}}_{\sigma^{\prime}},
\end{equation}
where $\xi_0$ is the orbital quantum dot energy,
$B=|\mu_B\mathbf{B}|$ represents the magnetic field splitting,
$\pmb{\tau}$ is a vector containing the Pauli spin-matrices,
  and $U$ is the on-site Coulomb interaction for
double occupancy. In a spin basis parallel to ${\bf B}$, the dot Hamiltonian
is diagonalized as
\begin{equation}  \label{HDd}
H_{D}^{{}}=\sum_{\tilde{\sigma}} (\xi_{0}^{{}}-\tilde{\sigma}
B)d^\dagger_{\tilde{\sigma}} d^{{}}_{\tilde{\sigma}} +U
n_{\tilde{\uparrow}}^{{}} n_{\tilde{\downarrow}}^{{}},
\end{equation}
where the $d_\sigma$ and $d_{\tilde{\sigma}}$ operators are
related by the unitary rotation
\begin{equation}  \label{dU}
d_\sigma=\sum_{\tilde{\sigma}} R_{\sigma\tilde{\sigma}}^{{}}
d_{\tilde{\sigma}},\quad \mathbf{R}=\left(
\begin{array}{cc}
\cos(\phi/2) & \sin(\phi/2) \\
-\sin(\phi/2) & \cos(\phi/2)
\end{array}
\right).
\end{equation}
Finally, the tunneling Hamiltonian is
\begin{equation}\begin{split}  \label{EqHT}
H_T^{{}}=&\sum_{\alpha=L,R}\sum_{k\sigma} \left(
t_{\alpha,k\sigma}^{{}}
c^\dagger_{\alpha,k\sigma}d^{{}}_{\sigma}+\mathrm{h.c.}
\right)\\
=&\sum_{\alpha=L,R}\sum_{k\sigma\tilde{\sigma}} \left(
t_{\alpha,k\sigma}^{{}}\mathbf{R}^{{}}_{\sigma\tilde{\sigma}}
c^\dagger_{\alpha,k\sigma}d^{{}}_{\tilde{\sigma}}+\mathrm{h.c.}
\right).
\end{split}\end{equation}
Here we allow for the tunneling matrix element
$t_{\alpha,k\sigma}$ to be spin-dependent, because the states in
the leads depend on the spin direction. Note that there is no
spin-flip associated with the tunneling here, i.e., there is no
spin-active interface, which would require the use of a
non-diagonal tunneling matrix $t_{\alpha,k\sigma\sigma'}$.
Depending on the parameters this would correspond to having an
angle between the lead magnetizations, which would modify the
details but not the general behavior that we discuss.

We define the energy-dependent coupling constants as
\begin{equation}  \label{Gammadef}
\Gamma_{\alpha}(\varepsilon)=2\pi\sum_{k\sigma}|t^{{}}_{\alpha,k\sigma}|^2\delta(
\varepsilon-
\xi_{\alpha,k\sigma}^{{}})=\sum_\sigma\Gamma_{\alpha,\sigma}(\varepsilon),
\end{equation}
and let $P_\alpha$ denote the polarization of the tunneling from
lead $\alpha$ defined through
$\Gamma_{\alpha,\sigma}(\varepsilon)=\frac{1}{2}\left(1+\sigma
P_\alpha^{{}}\right)\Gamma_\alpha(\varepsilon)$. Notice that
$P_{\alpha}\in [-1,1]$ such that $P_\alpha=\pm 1$ corresponds to
full spin-$\uparrow/\downarrow$ polarization and $P_\alpha=0$
corresponds to unpolarized leads. For parallel (antiparallel)
polarization of the leads the $P_\alpha$'s have the same
(opposite) sign.

In the basis where the dot part of the Hamiltonian is diagonal,
the coupling of the two dot states, $\tilde{\uparrow}$ and
$\tilde{\downarrow}$, to the lead $\alpha$ is given by a matrix in
the spin index, see also Eq.~(\ref{EqSigmaEll}),
\begin{equation}\label{EqGammaTilde}
\Gamma_{\alpha,\tilde{\sigma}\tilde{\sigma}'}(\varepsilon)=\frac{\Gamma_\alpha(\varepsilon)}{2}\times
\bigg\{\begin{array}{ll} \left(1+\tilde{\sigma}
P_\alpha\cos\phi\right)&\textrm{for }
\tilde{\sigma}=\tilde{\sigma}'\\
P_\alpha\sin\phi &\textrm{for }
\tilde{\sigma}\neq\tilde{\sigma}'
\end{array} .
\end{equation}

In the calculations using the DMRG and the density matrix
technique, we use a polarization of both leads less than $1$ for
technical reasons. The minority spin only introduces a smearing of
the results discussed for fully polarized leads.


\section{Linear response: DMRG} \label{SecDMRG}
\newcommand{\im}[1]{\mathrm{Im}\left[{#1}\right]}
\newcommand{\re}[1]{\mathrm{Re}\left[{#1}\right]}

\subsection{Tight-binding Hamiltonian} \label{40FAB: Chap: TB Hamiltonian}%
In order to apply the DMRG method to the model a discretized
version of the leads must be formulated. The simplest choice is to
model the leads as one-dimensional semi-infinite tight-binding
(TB) chains that are discretized appropriately. With this choice
and denoting the hopping matrix element between the resonant level
and the leads by $t_{\alpha,\sigma}$ ($\alpha=L,R$), the
tight-binding Hamiltonian reads $H^\mathrm{TB}=\sum_{\alpha=L,R}
H_{\alpha}^\mathrm{TB}+H_T^\mathrm{TB}+H_D$, where
\begin{eqnarray}%
H_{\alpha}^\mathrm{TB}  &=&-\sum_{n=2}^{\infty}\sum_\sigma
\frac{D}{2} \left(c^\dagger_{\alpha, n\sigma}
c^{\phantom\dag}_{\alpha, n-1\sigma}+c^\dagger_{\alpha, n-1\sigma}
c^{\phantom\dag}_{\alpha, n\sigma}\right),\nonumber\\~\\
H_T^\mathrm{TB} &=&
-\sum_{\alpha=L,R}\sum_\sigma\left(t_{\alpha,\sigma}
c^\dagger_{\alpha, 1
\sigma}d^{\phantom\dag}_{\sigma}+\mathrm{h.c.}\right),
\end{eqnarray}
and where $H_D$ is given in Eq.~(\ref{HD}). That is in the DMRG
implementation we work in the lead spin-basis and do not use the
diagonalized version of the dot part of the Hamiltonian. In
$H_{\alpha}^\mathrm{TB}$, $2D$ is the bandwidth of the
tight-binding chain representation of the leads corresponding to
the hopping amplitude $D/2$ between the internal sites in the
chains.

In order to link the different theoretical approaches applied to
solve the model, an expression for the effective energy-dependent
coupling constants between the single site and the tight-binding
leads, $\Gamma^\mathrm{TB}_{\alpha,\sigma}(\varepsilon)$, must be
established. For the one-dimensional tight-binding model of the
leads, these are given by \cite{DattaBook1995}
\begin{equation}\label{EqGammaTB}
  \Gamma_{\alpha,\sigma}^\mathrm{TB}(\varepsilon)=-2|t_{\alpha,\sigma}|^2~\im{
  g^r_{\alpha,\sigma}(1,1,\varepsilon)},
\end{equation}
where $g^r_{\alpha,\sigma}(1,1,\varepsilon)$ is the surface
component of the retarded Green function of the semi-infinite left
or right chain at energy $\varepsilon$. The surface of the
tight-binding chain is the first site, and the Green function
reads\cite{EconomouBook1983}
\begin{eqnarray}\label{EqChainGF}
    g^r_{\alpha,\sigma}(1,1,z) &=& 2\frac{z-\sqrt{z^2-D^2}}{D^2},
\end{eqnarray}
where $z=\varepsilon+i\eta$ is complex, and thus the imaginary
part of the Green function is finite only inside the band, $\pm
D$, and is proportional to the semi-elliptic density of states.
Thus the coupling constants are given by
\begin{equation}\label{EqTBdef}
\Gamma^\mathrm{TB}_{\alpha,\sigma}(\varepsilon)
=\frac{4|t_{\alpha,\sigma}|^2\sqrt{D^2-\varepsilon^2}}{D^2}.
\end{equation}
In Sec.~\ref{40FAB: Chap: Polarization} we discuss the
implementation of the polarization, and explain that half-filled
leads can be used, corresponding to $\varepsilon=0$.

\subsection{Modeling the polarization}\label{40FAB: Chap: Polarization} %
Full polarization of the leads is avoided for several reasons.
Most prominently, full polarization decouples one spin species in
a lead completely in the sense that the hopping matrix element
between the lead and the resonant level is zero for all angles.
Dealing with decoupled Hilbert spaces is undesired as it creates
numerical problems such as ill-conditioned matrices, making the
numerical solution of the resolvent equations
hard.\cite{Kuehner1999,Jeckelmann2002,Bohr2006}

Furthermore, there are single points where the model itself is
ill-defined for full polarization. At the angles $\phi=0$ and
$\phi=\pi$ the spin-flip process of the dot is inactive because of
the prefactor $\sin\phi$. Due to the full polarization also the
hopping matrix element for the minority spin connecting the lead
and the dot is zero. Thus the minority spin level is completely
decoupled and hence has a constant occupation. The occupation of
the majority spin level depends, however, on the occupation of the
minority spin level through the on-site Coulomb interaction term
$U n_{\uparrow} n_{\downarrow}$. That is the properties of the
model for these specific angles depend on the initial occupation
of the minority spin level, and no unique stationary state exists.

It should be noted that the qualitative behavior for (large)
partial
 and full polarization are similar except for the problem for
specific angles described above. It is, however, clear that a
decreased polarization in the leads tends to wash out the spin
dependence in the model, and in the limit of unpolarized leads all
spin characteristics are lost.\\

There is a certain freedom of choice in the modeling of the
polarization. Although the polarization is a property of the
leads, it can be modeled by spin-dependent hopping matrix elements
connecting the dot to the leads.\cite{PedersenPRB2005b} There are
different approaches to modeling the polarization of the leads and
we have chosen the simplest one to implement in the DMRG setup.
Rather than using spin-dependent \emph{filling} in the leads, we
use half-filled leads for both spin species, and model the
polarization by modifying the hopping matrix elements connecting
the leads and the dot. This is indicated in Fig.~\ref{FAB: DMRG
setup figure}, where we show the DMRG setup using a momentum-space
representation of the leads. This choice for the polarization
simplifies the DMRG setup significantly as \emph{identical
discretizations} can be used for the two spin species in each lead
such that the spin species are again treated equally apart from
the polarization dependent hopping matrix elements,
\begin{eqnarray}\label{EqDeft}
    t_{\alpha,\sigma} &=& t^0_{\alpha}\sqrt{\frac 12\big(1+\sigma P_{\alpha}\big)}.
\end{eqnarray}
In all calculations presented, we use identical polarizations of
the two leads, $P_L=P_R=P$, such that the coupling
to the leads are identical when $t^0_L=t^0_R$.\\

In the remainder of the article we measure all energies in units
of the sum of the coupling constants at the equilibrium chemical
potential, $\varepsilon=0$,
\begin{equation}
 \Gamma^0=\sum_{\alpha,\sigma}\Gamma_{\alpha,\sigma}^\mathrm{TB}(0).
\end{equation}
For the tight-binding chains this corresponds to measuring all
energies in units of
$\frac{4}{D}\left(|t_L^0|^2+|t_R^0|^2\right)$, see
Eqs.~(\ref{EqTBdef}),(\ref{EqDeft}).

\begin{figure}
\begin{center}
  \includegraphics[width=0.35\textwidth]{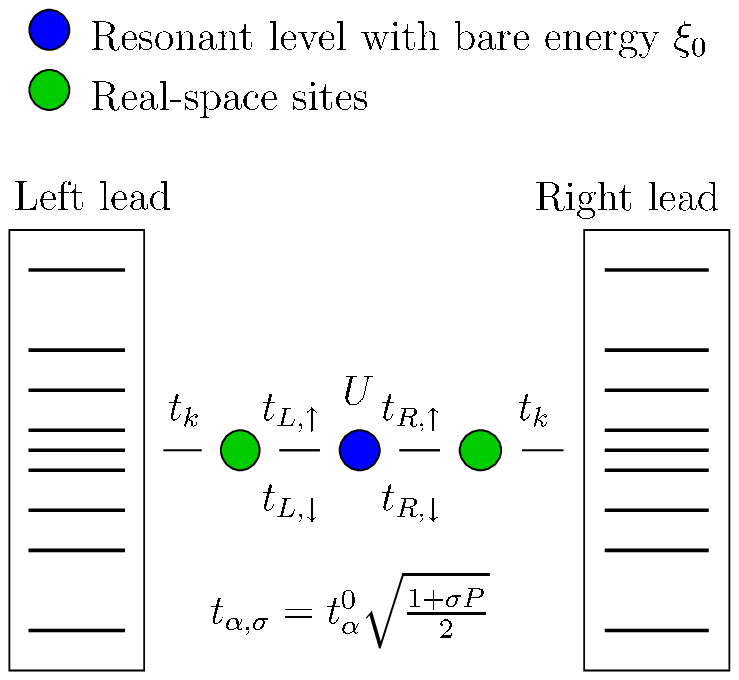}\\
\caption[DMRG setup for the FAB model.] {(Color online) Sketch of
the DMRG setup for the FAB model. Notice the implementation of the
polarization through the hopping matrix elements, where
$t_{\alpha,\sigma}$ and the on-site Coulomb interaction $U$ are
indicated in the figure. In the DMRG evaluation a single lead
mapping is used in combination with a momentum-space
representation of the single tight-binding lead.\cite{BohrPHD2007}
Here $t_k$ indicates a discretization dependent hopping to
\emph{all} states in the momentum-space. }\label{FAB: DMRG setup
figure}
\end{center}
\end{figure}

\subsection{Calculation of the conductance}
In order to obtain the conductance we make use of the
Meir-Wingreen formula\cite{Meir_Wingreen1992} rather than the Kubo
formula used in previous work.\cite{Bohr2006,Bohr2007} The
evaluation of the spectral function at zero bias and for
proportional couplings\cite{HaugJauho1996} can be done within a
single lead framework, effectively halving the lead size. For
parallel magnetizations of the leads, the FAB model falls in the
linear regime within this class of models, and thus the finite
size scaling for the evaluation of the spectral function is
significantly better than the evaluation of the Kubo formula,
enabling faster and more accurate calculations.\cite{BohrPHD2007}
Using DMRG we thus evaluate the two spin components of the full
spectral function in separate calculations, and therefore need to
recombine the spin resolved spectral functions into the total
conductance,
\begin{eqnarray}
  \lefteqn{G(\xi_0, \phi) =}\nonumber \\
  && \frac{e^2}{h} \sum_\sigma \frac{2|t_{L,\sigma}|^2 |t_{R,\sigma}|^2}
  {|t_{L,\sigma}|^2+|t_{R,\sigma}|^2} A_{\sigma}(\xi_0, \phi,\omega=0), \label{40FAB: MW for FAB}
\end{eqnarray}
where the polarization enters through the hopping matrix elements
$t_{\alpha, \sigma}$, and $A_\sigma$ denotes the spin resolved
spectral function of the dot. In this paper, we make the
assumption that the hopping matrix elements between the leads and
the dot are identical for both leads, $t^0_L=t^0_R$, and that the
polarizations in  both leads are identical, $P_L=P_R$, such that
$t_{L,\sigma}=t_{R,\sigma}=t_\sigma$.

In order to achieve the necessary precision in the DMRG
calculations a momentum-space representation of the leads is used.
Although the physics takes place at the Fermi level also energies
well away from the Fermi level need to be represented properly,
and the discretization scheme used should support this. We use a
discretization of the momentum part of each lead consisting of a
logarithmic discretization that covers a large energy span, and
switch to a linear discretization on the low-energy scale close to
the Fermi level.\cite{Bohr2007} All the DMRG calculations
presented in this paper were performed using $55$ sites in the
lead description, corresponding to $35$ sites scaled
logarithmically and $20$ sites scaled linearly around the Fermi
level.\cite{Bohr2007}

It should be noted that by virtue of the DMRG method \emph{all}
interactions are rigorously taken into account. The approximation
in the method presented lie in the use of a finite sized lead
which can be benchmarked in the non-interacting limit, and as such
the method used contains only controllable approximations.

\subsection{Results}\label{40FAB: Chap: Results} %
Using the momentum-space representation of the leads in the DMRG
setup we have calculated the spectral function, and using the
Meir-Wingreen formula in Eq.~\eqref{40FAB: MW for FAB} evaluated
the conductance. For different values of the magnetic field
strength $B$ and the on-site Coulomb interaction strength $U$, we
have calculated the conductance versus the angle $\phi$ between
the magnetic field and the polarization direction. In all
examples, we keep the polarization in the two leads identical,
$P_L=P_R=0.8$. As evident from the Hamiltonian, the model is
symmetric around $\phi=\pi$ since $\cos(2\pi-\phi)=\cos\phi$ and
$\sin(2\pi-\phi)=-\sin\phi$ such that only the spin-flip term in
$H_D$ acquires an insignificant phase [see Eq.~(\ref{HD})].
Therefore we confine our studies to angles in the interval
$\phi\in[0,\pi]$, and the interval $\phi\in[\pi,2\pi]$ is found by
reflecting the results around $\phi=\pi$.\\

In order to determine the discretization needed for the leads,
exact diagonalization calculations for the spectral function have
been performed and, using Eq.~\eqref{40FAB: MW for FAB}, compared
to the NEGF results in the non-interacting limit, $U=0$ (not
shown).\cite{BohrPHD2007}$^,$\footnote{For finite polarization
Eq.~(\ref{EgGFU0}) is not valid, but the results can easily be
found from Eqs.~(7)-(12) in Ref.~\onlinecite{PedersenPRB2005b}.}
By virtue of the exact diagonalization, the only error present in
this approach is the error due to the finite size of the leads.
The results show excellent agreement between the exact
diagonalization and the Green function results for a range of
parameter values, and we conclude that the modeling of the leads
is sufficient for resolving the model.

\begin{figure}
\begin{center}
\includegraphics[trim = 0mm 0mm 0mm 0mm,
clip,width=8cm]{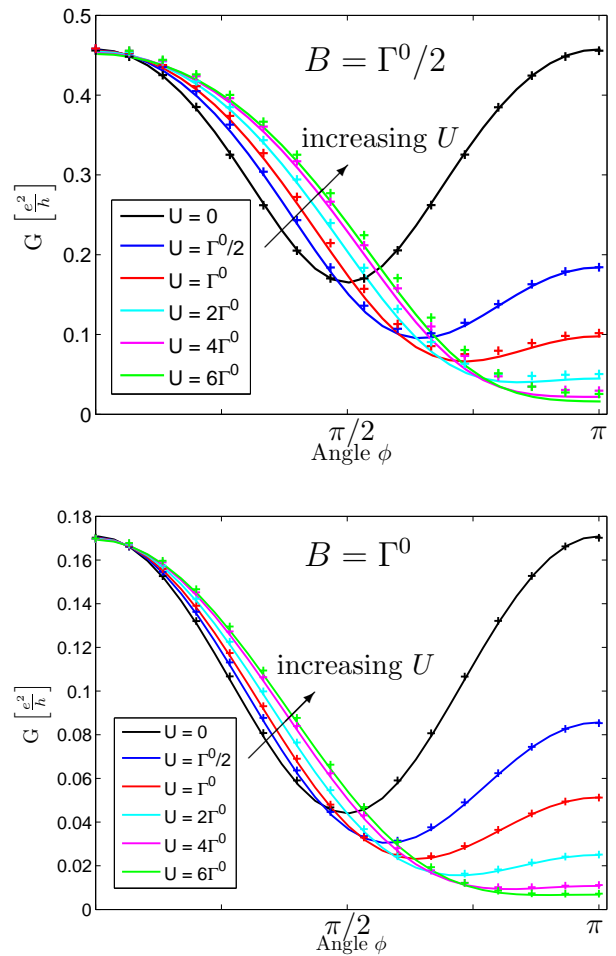} \caption{(Color online) The linear
conductance vs. the angle $\phi$ for $B=\Gamma^0/2$ and
$B=\Gamma^0$. The `+' symbols are the DMRG results and the full
lines are the non-equilibrium Green function results using the
Hartree-Fock approximation. We apply an elliptic density of
states, using Eq.~(\ref{EqTBdef}) for the coupling constant with
$D=2\Gamma^0$. The other parameters are $P=P_L=P_R=0.8$ and
$\xi_0=0$. The Green function result is exact in the
non-interacting limit, $U=0$, and is thus a rigorous benchmark for
DMRG in this limit. The good agreement demonstrates the accuracy
of the discretization, and confirms the capability of the DMRG.
Furthermore there is also a surprisingly good agreement between
these two methods for finite interactions as discussed in detail
in Sec.~\ref{SecMF}. } \label{FigLinResp}
\end{center}
\end{figure}

Having benchmarked the DMRG setup in the known limit of $U=0$ we
turn to the interesting regime of finite interactions. In
Fig.~\ref{FigLinResp} we show the results of the DMRG calculations
(`+' symbols), keeping the bare level resonant, $\xi_0=0$. The
calculations presented in each figure were performed keeping the
strength of the magnetic field $B$ fixed and varying the
interaction strength $U$ and the angle $\phi$, where the specific
parameter values are given in the plots.

For the parameter regimes considered here the numerical
zero-temperature DMRG results confirm the simple physical picture
sketched in the introduction. That is the linear conductance
versus the angle $\phi$ shows anti-resonances for $\phi=\pi/2$ in
the non-interacting limit, and a spin-valve behavior for strong
on-site interaction.

When $B/\Gamma^0$ decreases the maximum conductance increases as
the levels move closer to resonance, but the qualitative behavior
is the same when $U$ is varied. Previous work showed that for
fully polarized leads and $U=0$, the anti-resonances become
sharper for decreasing $B/\Gamma^0$ (see the left panel in Fig.~2
in Ref.~\onlinecite{PedersenPRB2005b}). Due to the finite
polarization, the sharpening of the anti-resonances is not very
clear in the DMRG results.

Finally we note that the cotunneling expression derived  under the
assumption $B\gg\Gamma^0$, see Eq.~(\ref{EqCotunPartial}),
reproduces the DMRG results fairly well already for $B=2\Gamma^0$
(not shown).


\section{Linear response: Mean-field solution}\label{SecMF}

Mean-field solutions are often problematic when considering
systems with only a few  degrees of freedom such as, e.g., transport
through quantum dots with only a few levels contributing to the
transport. In this case, the mean-field solution fails to describe
Coulomb blockade \cite{MuralidharanPRB2006} and can lead to
unphysical bistabilities at
finite bias due to a sudden switch between different transport
modes.

Surprisingly, Fig.~\ref{FigLinResp} shows that the mean-field
version of the Hamiltonian from Eq.~(\ref{Hmodel}) actually
reproduces the linear conductance results of the previous section
rather well, which we will explain below.

The mean-field version of the Hamiltonian in Eq.~(\ref{Hmodel}) is
obtained  by re-writing the interaction term as consisting of a
Hartree and a Fock-term
\begin{eqnarray}
H_\mathrm{Hartree} &=& U(d^\dag_\upTild d_\upTild
\ave{d^\dag_\downTild d_\downTild}+d^\dag_\downTild d_\downTild
\ave{d^\dag_\upTild d_\upTild}), \label{EqMF}\\ 
H_\mathrm{Fock} &=& -U(d^\dag_\upTild d_\downTild
\ave{d^\dag_\downTild d_\upTild}+d^\dag_\downTild d_\upTild
\ave{d^\dag_\upTild d_\downTild}),\nonumber
\end{eqnarray}
i.e., the replacement is $Un_\upTild n_\downTild \rightarrow
H_\mathrm{Hartree}+H_\mathrm{Fock}$. Keeping only the Hartree-term
gives a basis-dependent Hamiltonian, and for the FAB-model the two
spins are correlated (for $\phi\neq 0,\pi$) due to the coupling to
leads giving a non-vanishing Fock term.\footnote{For all figures
shown, we also did the calculations using only the Hartree-term.
Only minor difference occurred, which is caused by the fact that
$\ave{d^\dag_\upTild d_\downTild}$ is small due to the splitting
of the levels, $|\Eo-\En|\geq \Gamma^0$.}

From the mean-field Hamiltonian the linear conductance can be
obtained  using the non-equilibrium Green function formalism, and
the calculation is similar to the non-interacting calculation in
Ref.~\onlinecite{PedersenPRB2005b}. However here the
non-interacting dot Green function is replaced with the
Hartree-Fock dot Green function
\begin{equation}\label{EqGFHF}
\begin{split}
\mathbf{G}&_\mathrm{HF}^{-1}(\omega)=\\
&\left(%
\begin{array}{cc}
  \omega-(\xi_0- B + U\ave{d^\dag_\downTild d_\downTild}) & +U\ave{d^\dag_\downTild d_\upTild }  \\
  +U\ave{d^\dag_\upTild d_\downTild} & \omega-(\xi_0+ B + U\ave{d^\dag_\upTild d_\upTild}) \\
\end{array}%
\right).
\end{split}
\end{equation}
The generalized occupations have to be calculated
self-consistently through the relation
\begin{equation}\label{EqOccupations}
\ave{d^\dag_{\tilde{\sigma}} d_{\tilde{\sigma}^{\prime}}} =
\frac{-i}{2\pi}\int_{-\infty}^\infty d \omega
~G^<_{\tilde{\sigma}^{\prime}\tilde{\sigma}}(\omega),
\end{equation}
with the lesser Green function being
\begin{equation}\label{EqGless}
\mathbf{G}^<(\omega)=\imai f_L(\omega)
\mathbf{G}^r(\omega)\mathbf{\Gamma}_L\mathbf{G}^a(\omega)+ \imai
f_R(\omega)
\mathbf{G}^r(\omega)\mathbf{\Gamma}_R\mathbf{G}^a(\omega),
\end{equation}
where the expressions for the coupling constants,
$\mathbf{\Gamma}_\alpha$, are derived in App.~\ref{AppNEGF}.

The zero-temperature linear conductance is obtained
as\cite{HaugJauho1996,DattaBook1995}
\begin{equation}
G^\mathrm{MF}= \frac{e^2}{h}
\mathrm{Tr}\left[\mathbf{G}^a(0)\mathbf{\Gamma}_L\mathbf{G}^r(0)\mathbf{\Gamma}_R\right].
\end{equation}

\begin{figure}
  \includegraphics[width=0.9\columnwidth]{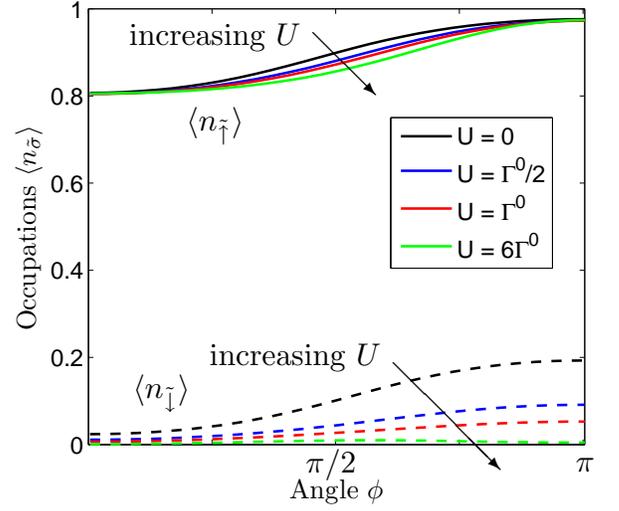}\\
  \caption{(Color online) The mean-field occupations vs. $\phi$ for $B=\Gamma^0/2$
  and four different values of the on-site Coulomb interaction.
The upper bunch of curves are for $\langle
n_{\tilde{\uparrow}}\rangle$ and the lower ones are for $\langle
n_{\tilde{\downarrow}}\rangle$.  The rest of the parameters are as
in Fig.~\ref{FigLinResp}.}\label{FigOccupationsMF}
\end{figure}

The results for $B=\Gamma^0/2$ and $B=\Gamma^0$ are shown in
Fig.~\ref{FigLinResp} together with the DMRG results. It is
clearly seen that for $B=\Gamma^0$ the results agree almost
exactly, but for $B<\Gamma^0$ deviations start to appear,
especially for angles around $\pi/2$. This rather surprising
success of the mean-field solutions can be understood by an
inspection of the occupations, see Fig.~\ref{FigOccupationsMF}:

First we notice that the couplings to the levels depend on the
angle $\phi$,  see Eq.~(\ref{EqGammaTilde}), and consider the
non-interacting case $U=0$. At $\phi=0$ the
spin-$\tilde{\uparrow}$ level is roughly $80\%$ occupied due to a
large broadening, but the spin-$\tilde{\downarrow}$ level is
almost unoccupied due to the large
$B/(\Gamma_{L,\tilde{\downarrow}\tilde{\downarrow}}+\Gamma_{R,\tilde{\downarrow}\tilde{\downarrow}})=5$.
At $\phi=\pi$ the spin-$\tilde{\uparrow}$-level is almost
decoupled and the occupation goes to $1$, whereas the occupation
of the spin-$\tilde{\downarrow}$ level is  $\sim 20\%$. For the
intermediate angles, there is a smooth cross-over between the two
regimes, see the black curves in Fig.~\ref{FigOccupationsMF}. When
calculating the occupations self-consistently for finite $U$, the
overall trend does not change: $\langle
n_{\tilde{\uparrow}}\rangle$ remains almost identical, and
$\langle n_{\tilde{\downarrow}}\rangle$ decreases for increasing
$U$, see Fig.~\ref{FigOccupationsMF}.

So the reason why the mean-field solutions performs relatively
well  (at least at zero temperature) is that when one of the
levels fluctuates the most, the other level has an occupation
being either approximately zero or $1$, i.e., the products of the
fluctuations vanishes.  For this specific model, the success of
the mean-field solution is due to the combination of the split
levels and the angle dependent couplings, which means that the
term $(n_{\tilde{\uparrow}}-\langle
n_{\tilde{\uparrow}}\rangle)(n_{\tilde{\downarrow}}-\langle
n_{\tilde{\downarrow}}\rangle)$, neglected in the mean-field
Hamiltonian, remains small for all angles. The largest deviation
between the mean-field results and the DMRG results is expected
for the angles around $\phi=\pi/2$, which is also observed in Fig.~\ref{FigLinResp}.\\

At elevated temperatures or for even smaller magnetic fields, the
mean-field solution is expected to perform worse due to larger
fluctuations of the occupations, but for finite temperatures no
exact results are currently available for comparison. Furthermore,
while we have shown here that the mean-field solutions are
somewhat fortuitously reliable to describe linear response, we
demonstrate in Sec.~\ref{SecFinitebias_Comp} that they fail for
the FAB model for finite bias.

\section{Linear response: Hubbard-I approximation}\label{SecHubbard1}

Another popular and widely used approximation is the Hubbard-I
approximation (HIA) which corresponds to a decoupling of equations
of motion for the Green's functions at a higher level of the
hierarchy than the simple mean-field approximation used in the
previous section. Therefore, one could expect better performance
of this approximation compared to the mean-field approximation.
This expectation is further supported by the fact that HIA
naturally arises as the lowest-order approximation in the theories
developing perturbation series around the state of the isolated
system, i.e.~around the so-called ``atomic limit". Despite the
lack of existence of the standard Wick theorem due to the
non-Gaussian nature of the unperturbed system with arbitrary
correlations, systematic (even renormalized) perturbation theory
reportedly exists \cite{SandalovIJQC2003,Ovchinnikov2004} and HIA
is in a certain sense the lowest order in that expansion. Indeed,
it turns out that HIA can rather simply and yet correctly describe
the non-trivial effects of single-level broadening in the Coulomb
blockade regime of transport studied by other sophisticated
methods\cite{PedersenPRB2005a, KoenemannPRB2006} and confirmed
experimentally.\cite{KoenemannPRB2006} Therefore HIA appears to be
the solution to the problem of a simultaneous description of
interference/broadening and interaction.

However, this optimistic picture breaks down as soon as more
complicated systems are addressed, namely, any system where
coherence between different transport channels is involved. The
FAB model is one such example and we will demonstrate the
breakdown of HIA for this model. A similar situation arises for
the problem of a single spin-degenerate level with local Coulomb
interaction coupled to superconducting leads in the so-called
$\pi$-junction regime, where the Josephson supercurrent is
observed. The HIA is known to fail to predict this theoretically
and experimentally well-established fact, see
e.g.~Ref.~\onlinecite{OsawaCM2008}, Fig.~4a. The situation in the
case of the FAB model is even more severe since the HIA failure is
not just physical, as in the superconducting case, but the results
are even {\em mathematically inconsistent}. Fundamental analytical
identities such as $\left( \mathbf{G}^{r}\right)^{\dagger
}=\mathbf{G}^{a}$ and, consequently, the hermiticity of the
density matrix is broken within the HIA.

To demonstrate this explicitly, we use the Hamiltonian in
Eq.~(\ref{Hmodel}) and perform analogous derivations as in
Refs.~\onlinecite{RudzinskiPRB2005}, \onlinecite{FranssonPRB2005}
for simpler models without any magnetization at all or without the
local splitting, respectively. We arrive at the matrix equation
\begin{equation}
\mathbf{G}(\varepsilon)=\mathbf{g}(\varepsilon)
+\mathbf{g}(\varepsilon)\mathbf{\Sigma}(\varepsilon)\mathbf{G}(\varepsilon)
\label{EqG_versus_g}
\end{equation}%
for the causal Green function
$G_{\tilde{\sigma}\tilde{\sigma}'}(\varepsilon)$ (in energy
representation) of the central dot. Here
\begin{equation}
\Sigma _{\tilde{\sigma} \tilde{\sigma} ^{\prime}}(\varepsilon)
=\sum_{\alpha k\sigma}|t_{\alpha, k\sigma}^{{}} |^{2}
\left(\mathbf{R}^{\dagger}\right)_{\tilde{\sigma} \sigma}^{{}}\frac{1}{%
\varepsilon-\xi_{\alpha,k\sigma}^{{}}}\mathbf{R}_{\sigma\tilde{\sigma}^{\prime}}^{{}}.
\label{EqSigma}
\end{equation}%
is the self-energy matrix due to the coupling to the contacts, and
we introduced the auxiliary Green function
\begin{equation}
\mathbf{g}(\varepsilon)=\left(
\begin{array}{cc}
\frac{\varepsilon -\xi _{\upTild }
-U(1-\langle d^\dag_{\downTild}d_{\downTild}\rangle )}
{(\varepsilon -\xi _{\upTild })(\varepsilon-\xi _{\upTild }-U)}
& \frac{-U\langle d^\dag_{\downTild}d_{\upTild}\rangle }
{(\varepsilon -\xi _{\upTild })(\varepsilon -\xi_{\upTild }-U)} \\
\frac{-U\langle d^\dag_{\upTild}d_{\downTild}\rangle }{(\varepsilon -\xi
_{\downTild })(\varepsilon -\xi _{\downTild }-U)} &
\frac{\varepsilon -\xi _{\downTild }
-U(1-\langle d^\dag_{\upTild}d_{\upTild}\rangle)}
{(\varepsilon -\xi_{\downTild})(\varepsilon -\xi _{\downTild }-U)}%
\end{array}%
\right) \mathbf{,}  \label{Eqg}
\end{equation}%
where $\xi_{\upTild/\downTild}=\xi_0\mp B$ are the level energies
in the basis parallel to ${\bf B}$ as given in Eq.~(\ref{HDd}).
For vanishing splitting, $B=0$, these are precisely the
Eqs.~(15)--(21) of Ref.~\onlinecite{RudzinskiPRB2005}. The
retarded and advanced Green function are then again obtained by
${\bf G}^{r/a}(\varepsilon)={\bf G}(\varepsilon\pm\imai 0^+ )$,
respectively. In addition, we define the retarded and advanced
components of the other functions in the same way.

A general condition relating the retarded and advanced Green's
functions is $\left( \mathbf{G}^{r}\right)^{\dagger
}=\mathbf{G}^{a}$. As Eq.~(\ref{EqG_versus_g}) implies
\begin{equation}
\mathbf{g}^{r/a}=\left[ \left( \mathbf{G}^{r/a}\right)
  ^{-1}+\mathbf{\Sigma }^{r/a}\right] ^{-1},
\end{equation}
and $\left( \mathbf{\Sigma}^{r}\right)^{\dagger
}=\mathbf{\Sigma}^{a}$ from Eq.~\eqref{EqSigma} this requires
$\left( \mathbf{g}^{r}\right) ^{\dagger }=\mathbf{g}^{a}$. This
is, however, only compatible with Eq.~(\ref{Eqg}) if the
off-diagonal elements are mutually complex conjugated. For
non-zero coherences $\langle d^\dag_{\downTild}d_{\upTild}\rangle
=\langle d^\dag_{\upTild}d_{\downTild}\rangle^*\neq0$ and finite
level splitting $\xi _{\upTild}\neq\xi_{\downTild}$ pertinent to
the FAB model this condition is not satisfied and, thus, HIA
breaks the necessary mathematical condition for the Green's
functions. Similarly, self-consistent evaluation of the
(generalized) populations in the spirit of
Eq.~\eqref{EqOccupations} in thermal equilibrium (since we study
currently the linear response only) would yield a non-hermitian
density matrix, yet another mathematical problem stemming from the
inconsistency of the HIA equations.

It should be stressed that the inconsistency only shows up if both
the local level splitting {\em and} the coherences between
different spin states (stemming from non-collinear magnetization
arrangement) are present. Therefore, the inconsistency has
apparently not been noticed
before\cite{RudzinskiPRB2005,FranssonPRB2005} since previously
employed models do not contain both necessary ingredients.

Nevertheless, the FAB model is both mathematically and physically
a realistic model and the failure of HIA reveals problems inherent
in that approximation. As mentioned before, the HIA is in some
sense the lowest-order expansion in reportedly systematic theories
based on Hubbard operator Green's
functions.\cite{SandalovIJQC2003} One can show that the same
problems carry over to the many-body formalism, see
Ref.~\onlinecite{FranssonPRB2005} for a pedagogical overview of
the connection between the formulation of HIA in the standard as
well as the many-body formalism.


\section{Finite bias: 2vN approach}\label{Sec2vN}

So far we have only considered linear response and zero
temperature. In this section we apply a density matrix formalism
developed in Ref.~\onlinecite{PedersenPRB2005a}, giving access to
the regime of finite bias and finite temperature. The method works
in a basis of many-particle eigenstates for the dot Hamiltonian,
thereby including all interactions on the dot exactly. Correlated
transitions between the lead and the dot states with up to two
different electron states are included exactly, which suggest the
notation second order von Neumann approach (2vN). By solving the
resulting set of equations for the steady-state, a certain class
of higher-order processes are also included. Interference effects
are also included by a full treatment of the nondiagonal density
matrix elements.

The dot part of the Hamiltonian, $H_D$, has four many-particle
 eigenstates
$\{|0\rangle,|\tilde{\uparrow}\rangle,|\tilde{\downarrow}\rangle,|2\rangle\}$,
where $|2\rangle=\tilde{d}^\dag_\downarrow
\tilde{d}^\dag_\uparrow|0\rangle$, and the energies are $0, \Eo =
\xi_0-B , \En = \xi_0+B$ and $\Eto=\Eo+\En+U$, respectively.
Inserting a complete set of dot states in the tunneling
Hamiltonian from Eq.~(\ref{EqHT}) gives
\begin{equation}
\begin{split}
H_T=\sum_{k\sigma\alpha,ab}\left(T_{ba}(k\sigma\alpha)|b\rangle
\langle a|c_{\alpha,k\sigma} +c^{\dag}_{\alpha,k\sigma} |a\rangle
\langle b|T_{ba}^*(k\sigma \alpha)\right)
\end{split}\end{equation}
where $a,b$ denotes the dot many-particle eigenstates and
$T_{ba}(k\sigma\alpha)=\sum_\mu
t_{\alpha,k\sigma}^*R_{\sigma\mu}\langle b|d_\mu^\dagger|a\rangle$
are the couplings between these states and the lead states.

Inserting these coupling matrix elements and eigenenergies in
Eqs.~(10),(11) in Ref.~\onlinecite{PedersenPRB2005a} gives a
closed set of equations for the elements of the reduced density
matrix, which can be solved numerically. As for the DMRG
calculation we assume the leads to be $80\%$ polarized.
Significantly larger polarizations ($\approx 1$) are difficult to
handle numerically, especially in the linear conductance regime.
This also limits the values for the bias voltages and temperatures
which can be used.

For the finite bias calculations presented here, we use
a constant density of states, so that effects due to the change of
the chemical potentials are not superimposed by changes in the
contact couplings. For numerical purposes, we
implement a finite
band width with elliptically shaped edges at $0.95
D<|\varepsilon|<D$, where $D$ surpasses all other relevant energy
scales. In all plots, the bias voltage $V$ is
applied symmetrically, $\mu_L=-\mu_R=eV/2$.

Before considering finite bias, we have carefully inspected the
results for low bias ($eV/\Gamma^0=0.05-0.1$) and low temperature
($k_BT/\Gamma^0=0.05-0.1$) for $B/\Gamma^0=0.5, 1, 2$ for all
angles $\phi$ (not shown). For the non-interacting case, the exact
NEGF results are reproduced for all parameters tested. From the
low-bias results we have extracted a numerical value for the
linear conductance, and the results show almost quantitative
agreement with the exact DMRG results for all tested values of $U$
and all angles. The discrepancies can be attributed to the (small
but) finite bias and the finite  temperature used in the density
matrix calculation. We conclude that the 2vN approach is capable
of describing the effect of the interactions and the coherence in
the low-bias regime for the model system considered.

\begin{figure}
  \centering
  \includegraphics[trim = 0mm 0mm 0mm 0mm, clip, width=0.38\textwidth]{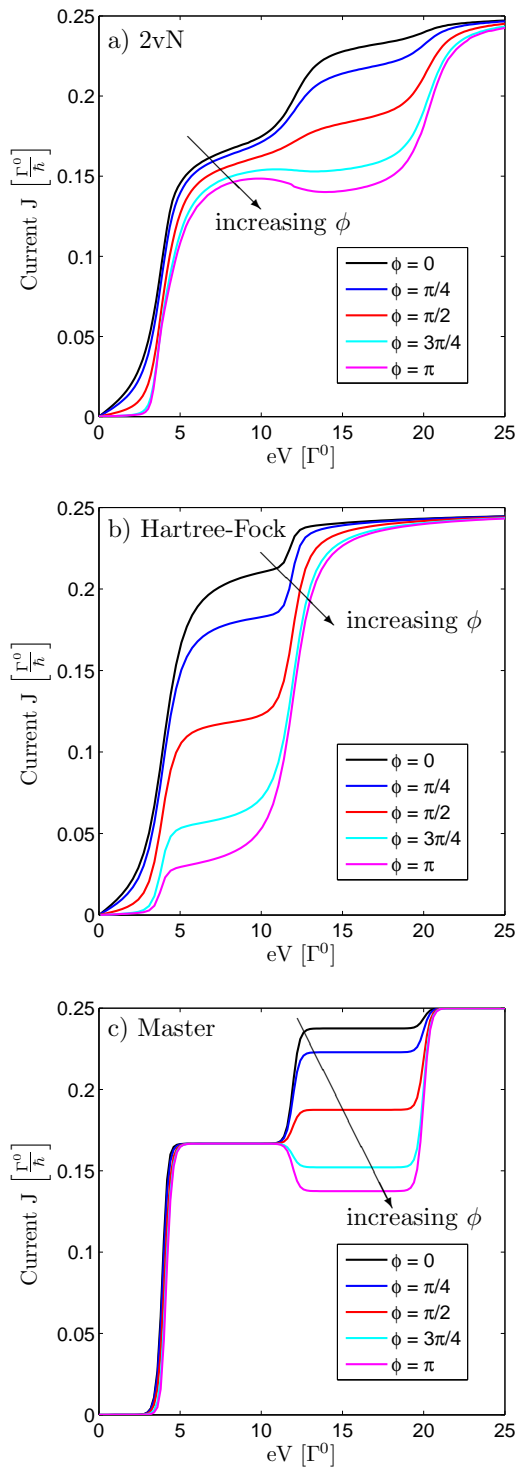}\\
  \caption{(Color online) The current versus bias voltage for five different angles $\phi$
    obtained using the 2vN density matrix formalism (a), the mean-field
    Hartree-Fock approach (b), and the master equation approach (c).
    The parameters are: $\xi_0=0$, $B=2\Gamma^0$, $U=8\Gamma^0$,
    $k_BT=0.1 \Gamma^0$, $P_L=P_R=0.8$ and the bias is applied symmetrically.
    For the 2vN-method we used a constant density of states with a half
    band-width $D=20\Gamma^0$, see the main text, while for the Hartree-Fock
    calculation the wide-band limit is applied, i.e.
    the real-part of the self-energy was neglected. }\label{FigFiniteBias}
\end{figure}

Fig.~\ref{FigFiniteBias}a) shows the current versus bias voltage
for different angles, where the bare level is on resonance,
$\xi_0=0$, $B=2\Gamma^0$ and an on-site Coulomb interaction
$U=8\Gamma^0$. Shoulders in the current are expected if half the
bias matches the single-electron transition energies in the dot.
This happens at $eV/\Gamma^0=4$ for the transitions
$\ket{0}\rightarrow\ket{\tilde{\uparrow}}$ and
$\ket{0}\rightarrow\ket{\tilde{\downarrow}}$, at $eV/\Gamma^0=12$
for $\ket{\tilde{\downarrow}}\rightarrow\ket{2}$, and finally at
$eV/\Gamma^0=20$ for $\ket{\tilde{\uparrow}}\rightarrow\ket{2}$.

In the low-bias regime, $eV/\Gamma^0<4$, the current is suppressed
when the angle $\phi$ is increased from $0$ to $\pi$ due to the
spin-valve effect (see also, e.g., Fig.~\ref{FigLinResp} for large
$U/\Gamma^0$).

In the intermediate regime $12<eV/\Gamma^0<20$, the current shows
a very pronounced angular dependence, with a significant current
drop between $\phi=0$ and $\phi=\pi$. For $\phi\approx \pi$ one
even detects negative differential conductance around
$eV=12\Gamma^0$. In this region, the lower spin level
$\tilde{\uparrow}$ can be filled not only by the process
$\ket{0}\rightarrow\ket{\tilde{\uparrow}}$ but also by
$\ket{\tilde{\downarrow}}\rightarrow\ket{2}$. It is thus more
likely to be filled compared to lower biases $eV<12\Gamma^0$.  In
the case  $\phi=0$, $\tilde{\uparrow}$ is aligned with the lead
polarization and therefore exhibits high tunneling rates. Thus its
increased occupation probability goes along with an increase of
current around $eV=12\Gamma^0$. In contrast, for $\phi=\pi$,
$\tilde{\uparrow}$ is pointing against the lead polarization and
therefore has a low tunneling rate, explaining the drop of current
around $eV=12\Gamma^0$.  For intermediate angles there is a smooth
cross-over between the two limits. Here the non-diagonal elements
of the density matrix are non-vanishing and quantum coherence
plays a role, as the two dot states are superpositions of the lead
spins. Off-diagonal elements are also important to include in
transport through dots coupled to non-collinear ferromagnetic
leads, even in the absence of an applied magnetic
field.\cite{BraunPRB2004}$^{,}$\footnote{The angular dependence in
presence of spin-split levels is also discussed for one polarized
and one unpolarized lead in Ref.~\onlinecite{DattaCM2006}.}

\section{Finite bias: mean-field NEGF and master equation}\label{SecFinitebias_Comp}
Figure~\ref{FigFiniteBias}b) shows the
current versus applied bias calculated using the Hartree-Fock
mean-field version of the Hamiltonian, see
Eqs.~(\ref{EqMF})-(\ref{EqGless}) and App.~\ref{AppNEGF}, within
the NEGF formalism using a self-consistent calculation of the
occupations, see, e.g., Ref.~\onlinecite{DattaBook1995}. The
parameters are identical to Fig.~\ref{FigFiniteBias}a) in order to
allow for a direct comparison.

In the low-bias regime ($eV<4\Gamma^0$), where the average
occupations $\langle n_{\tilde{\uparrow}}\rangle$, $\langle
n_{\tilde{\downarrow}}\rangle$   are close to one and zero,
respectively, the results agree with the 2vN formalism of
Sec.~\ref{Sec2vN}.
This goes well with the observation from section \ref{SecMF}, that
the conductance is well reproduced within the mean-field model.

In the region $4<eV/\Gamma^0<12$, we find $\langle
n_{\tilde{\uparrow}}\rangle\approx 0.5$ as the lower energy state
$\tilde{\uparrow}$ is in the window between the Fermi levels with
symmetric coupling to both contacts. Thus the higher energy state
$\tilde{\downarrow}$ has the energy $B+U/2=6\Gamma^0$ which is
above the emitter Fermi level and $\langle
n_{\tilde{\downarrow}}\rangle\approx 0$. If $\tilde{\uparrow}$ is
aligned with the lead polarization (i.e. $\phi=0$), the current is
larger, while it is low for $\phi=\pi$ as seen in
Fig.~\ref{FigFiniteBias}b). Finally, at $eV/\Gamma^0>12$, the
upper level $\tilde{\downarrow}$ enters the  window between the
Fermi levels, takes an average occupation  $\langle
n_{\tilde{\downarrow}}\rangle\approx 0.5$, and contributes also to
the current. The repulsion of the levels by $U/2$ is an artefact
of the mean-field model, and correspondingly neither the current
values nor the shoulders agree with the more detailed 2vN results
shown in Fig.~\ref{FigFiniteBias}a).

Fig.~\ref{FigFiniteBias}c) shows the corresponding result from the
master equation approach
\cite{KinaretPRB1992,PfannkuchePRL1995,MuralidharanPRB2006} where
the occupations of the many-particle states are determined by
electron hopping processes to and from the leads. While this
approach does not provide any current for low biases, where
cotunneling dominates the transport, it provides reliable results
for the current plateaus. In particular, the occurrence of
negative differential conductance for $\phi\approx \pi$ for biases
around $12 \Gamma^0/e$ is confirmed. Note that the presence of
pronounced steps of width $k_BT$ is due to the entire neglect of
broadening in this approach. Similar results are obtained by
taking into account nondiagonal density matrices within the 1vN
approach \cite{PedersenPRB2007} (not shown).

\section{Conclusion}\label{SecConclusion}

In this article, we provided a full description of the
Ferromagnetic Anderson model with applied magnetic field $B$
(FAB). We have successfully implemented the density matrix
renormalization group (DMRG) method, which provides the linear
conductance for arbitrary strength of the on-site Coulomb
interaction and arbitrary level splitting. The data interpolate
between the known results of non-equilibrium Green functions
(NEGF) for zero interaction and the cotunneling results for large
level splitting. A key result is the strong suppression of
conductance with increasing on-site Coulomb interaction if the
magnetic field on the dot is opposite to the lead polarization.

The DMRG results can serve as a benchmark for different
approaches, where we find that both the second order von Neumann
(2vN) approach and the NEGF approach with mean-field interaction
give reliable results for the conductance. While the 2vN approach
also provides plausible results for finite bias, the mean-field
NEGF fails due to the wrong treatment of partially occupied
states.

For finite bias the 2vN approach predicts a strong dependence of
the current on the direction of the magnetic field in the
intermediate bias region. Here negative differential conductance
is predicted if the magnetic field on the dot is opposite to the
lead polarization. This feature can also be qualitatively obtained
from the simpler master equation approach.

Finally we have shown that the Hubbard I approximation leads to
unphysical results for this particular model. This shows that the
FAB model constitutes a sensitive test case for different
approaches due to its involved interplay between interference,
broadening, and interaction.

\acknowledgments J.~N.\ and A.~W.\ acknowledge support by the
Swedish Research Council (VR). Parts of this work were done while
D.~B.\ was a student at the Department of Micro and
Nanotechnology, Technical University of Denmark. D.~B.~further
acknowledges support from the HPC-EUROPA under project
RII3-CT-2003-506079, supported by the European Commission. The
DMRG calculations were performed on the HP XC4000 at the Steinbuch
Center for Computing (SCC) Karlsruhe under project RT-DMRG, with
support through project B2.10 of the DFG -- Center for Functional
Nanostructures.  The work of T.~N.\ is a part of the research plan
MSM 0021620834 financed by the Ministry of Education of the Czech
Republic and was also partly supported by the grant number
202/08/0361 of the Czech Science Foundation. Finally, K.F.
acknowledged the European Community's Seventh Framework Programme
(FP7/2007-2013) under grant `SINGLE' no 213609.


\appendix

\section{Cotunneling expression}\label{AppCotun}
The expression for the cotunneling current in
Eq.~(\ref{EqCotunFull}) can easily be generalized to arbitrary
polarization and off-resonant transport, $\xi_0\neq 0$. For
identical polarizations of the leads, $P_L=P_R=P$, it reads
\begin{widetext}
\begin{equation}\label{EqCotunPartial}
\begin{split}
G^\mathrm{cotun}=&\frac{e^2\Gamma_L^0\Gamma_L^0}{2\pi \hbar}
\Bigg[
\left(\frac{1+P}{2}\right)^2\left(\frac{\cos^2(\phi/2)}{\xi_0-B}+
\frac{\sin^2(\phi/2)}{\xi_0+B+U} \right)^2\\
& + 2
\frac{(1+P)(1-P)}{4}\left(\frac{-\sin(\phi/2)\cos(\phi/2)}{\xi_0-B}+
\frac{\sin(\phi/2)\cos(\phi/2)}{\xi_0+B+U} \right)^2
+\left(\frac{1-P}{2}\right)^2\left(\frac{\sin^2(\phi/2)}{\xi_0-B}+
\frac{\cos^2(\phi/2)}{\xi_0+B+U} \right)^2\Bigg].
\end{split}
\end{equation}
\end{widetext}

\section{Equations for the NEGF solution}\label{AppNEGF}
For completeness we present here  the equations for the
non-equilibrium Green function calculations within the
Hartree-Fock approximation. Using the equation-of-motion
technique, the Green functions in the diagonal basis
are\cite{PedersenPRB2005b}
\begin{subequations}
\begin{align} \label{Gr0}
\mathbf{G}^{r,a}(\varepsilon)&=\left(\mathbf{G}_\mathrm{HF}^{-1}-\mathbf{\Sigma}^{r,a}_L
(\varepsilon)-\mathbf{\Sigma}^{r,a}_R(\varepsilon)\right)^{-1}, \\
[\mathbf{\Sigma}^{r,a}_\alpha
(\varepsilon)]_{\tilde{\sigma}\tilde{\sigma}^{\prime}}&=\sum_{k\sigma}
(\mathbf{R}^\dagger)_{\tilde{\sigma}\sigma}
|t^{{}}_{\alpha,k\sigma}|^2 g^{r,a}_{\alpha,
k\sigma}(\varepsilon)\mathbf{R}_{\sigma
\tilde{\sigma}^{\prime}}^{{}}\,,
\end{align}
\end{subequations}
where $g^{r,a}_{\alpha,
k\sigma}(\varepsilon)=(\varepsilon-\xi_{\alpha,k\sigma}^{{}}\pm
i0^+)^{-1}$ and
$\mathbf{\Gamma}_{\alpha}=-i\left[\mathbf{\Sigma}^{r}_\alpha-\mathbf{\Sigma}
^{a}_\alpha\right]$. The Green function
$\mathbf{G}_\mathrm{HF}^{-1}(\varepsilon)$ is stated in
Eq.~(\ref{EqGFHF}).  %

For the tight-binding chain of Sec.~\ref{SecDMRG} with elliptic bands we find
\begin{equation}\begin{split}
\Gamma_\alpha(\varepsilon) \equiv&
2\pi\sum_{k\sigma}|t_{\alpha,k\sigma}|^2\delta(\varepsilon-\xi_{\alpha,k\sigma})\\
=&\sum_\sigma \Gamma^\mathrm{TB}_{\alpha,\sigma}(\varepsilon)
=\frac{4|t_{\alpha}^0|^2\sqrt{D^2-\varepsilon^2}}{D^2}\, ,
\end{split}\end{equation}
where Eqs.~(\ref{EqTBdef}), (\ref{EqDeft}) have been used. The
$\upTild\upTild$-component of the self-energy becomes
\begin{equation}\begin{split}
\left[\Sigma^{r,a}_\alpha(\varepsilon)\right]_{\upTild\upTild}=&
\sum_{k\sigma} \frac{(\mathbf{R}^\dagger)_{\upTild\sigma}
|t^{{}}_{\alpha,k\sigma}|^2 \mathbf{R}_{\sigma \upTild}^{{}}}
{\varepsilon-\xi_{\alpha,k\sigma}\pm i 0^+}\\
=&\frac{1}{2}(1+P_\alpha\cos\phi)\int
\frac{d\varepsilon^{\prime}}{2\pi}
\frac{\Gamma_{\alpha}(\varepsilon')} {\varepsilon-\varepsilon' \pm
i 0^+}\, ,
\end{split}\end{equation}
where we used the definition of the polarization
[see below Eq.~(\ref{Gammadef})].

The principal part of the integral can be found analytically for
the elliptic density of states
\begin{equation}
\mathcal{P}\int^{D}_{-D}
\frac{d\varepsilon'}{2\pi}\frac{\Gamma_{\alpha}(\varepsilon')}
{\varepsilon-\varepsilon'}
=\frac{\Gamma_\alpha(0)\varepsilon}{2D}~~\mathrm{for}~|\varepsilon|<D,
\end{equation}
where it has been used that $\Gamma_\alpha(0)=4|t^0_\alpha|^2/D$.

The other components are calculated similarly, and the full expression for the self-energy becomes%
\begin{multline}\label{EqSigmaEll}
\Sigma^{r,a}_{\alpha}(\varepsilon)=\frac{1}{2}
\left(%
\begin{array}{cc}
  1+P_\alpha\cos\phi & P_\alpha\sin\phi \\
  P_\alpha\sin\phi & 1-P_\alpha\cos\phi \\
\end{array}%
\right)\\
\times\left[\frac{\varepsilon\Gamma_{\alpha}(0)}{2D}\mp
\frac{i\Gamma_\alpha(\varepsilon)}{2}\right].
\end{multline}

After a self-consistent evaluation of the Green functions, the
current can within the mean-field approximation be evaluated  as
\begin{equation}
\begin{split}
J^\mathrm{MF}=\frac{1}{2\pi\hbar}\int_{-\infty}^\infty
d\varepsilon
\mathrm{Tr}&\left[\mathbf{G}^a(\varepsilon)\mathbf{\Gamma}_L(\varepsilon)
\mathbf{G}^r(\varepsilon)\mathbf{\Gamma}_R(\varepsilon)\right]\\
&~\times \left[f_L(\varepsilon)-f_R(\varepsilon)\right],
\end{split}
\end{equation}
with
$f_\alpha(\varepsilon)=1/\left[e^{(\varepsilon-\mu_\alpha)/k_BT}+1
\right]$, $\alpha=L,R$, being the Fermi function.


\bibliographystyle{apsrev}

\end{document}